\documentclass{iaus}
\usepackage{graphicx}

\title[Bulges in CDM] 
{The hierarchical build-up of bulges in CDM}

\author[S. Khochfar]   
{Sadegh Khochfar}

\affiliation{Sub-Department of Astrophysics, University of Oxford,
Denys Wilkinson Bldg., Keble Road, OX1 3RH, Oxford, UK, \break email: sadeghk@astro.ox.ac.uk\\[\affilskip]}

\pubyear{2004}
\volume{xxx}  
\pagerange{119--126}
\date{?? and in revised form ??}
\jname{Proceedings Title IAU Symposium}
\editors{A.C. Editor, B.D. Editor \& C.E. Editor, eds.}
\begin{document}

\maketitle

\begin{abstract}
We investigate the hierarchical build-up of stars in bulges within the 
standard $\Lambda$-cold dark matter scenario. By separating the population 
into stars born during starbursts that accompany the formation of 
spheroids in major mergers ({\it starburst} component), and stars that are 
previously formed in discs of progenitor galaxies ({\it quiescent} component)
and added to the spheroid by dynamical interaction. Our results are summarised
as follows: bulges that form early have larger starburst fraction and hence 
should be smaller than their counter parts that form later. The quiescent 
fraction in bulges is an increasing function of bulge mass, becoming constant 
at $M_{\rm{q}}/M_{\rm{bul}} \sim 0.8$, mainly due to the infall of satellite 
galaxies that contribute disc stars to the bulge. Minor mergers 
are an order of magnitude more frequent than major mergers and must play a 
significant role in the evolution of bulges. Above the critical mass 
$M_{\rm{c}}\sim 3 \times 10^{10}$ M$_{\odot}$ most of the stars in the 
universe are in spheroids, which at high redshift are exclusively elliptical 
galaxies and at low redshifts partly bulges. Due to the enhanced evolution of 
galaxies ending up in high density environments, the starburst fraction and 
the surface mass densities of bulges below $M_{\rm{c}}$  should be enhanced
with respect to field galaxies. Dissipation during the formation of massive 
bulges in present day early-type spirals is less important than for the 
formation of present day elliptical galaxies of the same mass thereby 
explaining the possible difference in phase-space densities between spiral 
galaxies and elliptical galaxies.

\end{abstract}

\firstsection 

\section{Introduction}\label{intro}

The close resemblance of elliptical galaxies and {\it classical} bulges has 
lead to the widely accepted assumption that they have the same origin. 
Profiles of elliptical galaxies and bulges are nicely fit by  
Sersic-laws. The fact that super-massive black holes in bulges 
also follow the fundamental $M_{\bullet}$-$\sigma$-relation (\cite[Sarzi et al. 2001]{2001ApJ...550...65S}) 
provides further evidence for a common formation scenario of elliptical  
galaxies and classical bulges.

Early work by \cite{tt72} showed that elliptical galaxies can be the result of
a major merger between two spiral galaxies. 
Subsequent numerical simulations showed that indeed various properties of
elliptical galaxies and classical bulges can be recovered from simulations 
that use cosmological self-consistent initial orbital parameters (\cite[Khochfar \& Burkert 2006]{kb06}) 
for merging systems (see e.g. \cite[Barnes \& Hernquist 1992]{ba92}; 
\cite[Naab \& Burkert 2003]{nb03}; \cite[Jesseit, Naab \& Burkert 2005]{2005MNRAS.360.1185J}; \cite[Naab, Jesseit \& Burkert 2006]{2006MNRAS.372..839N}; 
\cite[Jesseit et al. 2007]{2007MNRAS.376..997J}). 
As a consequence it should be possible to generalise results for 
the formation of elliptical galaxies to the formation of classical bulges 
and to speak more general of the formation of spheroids 
(\cite[Khochfar \& Silk 2006a]{2006MNRAS.370..902K}).
 E.g. it has been predicted that massive spheroids
form in dry major mergers of elliptical galaxies, and that intermediate 
mass spheroids form as a result of a major merger between an elliptical and a 
spiral galaxy (\cite[Khochfar \& Burkert 2003]{kb03}; 
\cite[Naab, Khochfar \& Burkert 2006]{nk06}).  \cite{2006MNRAS.370..902K} find 
that this is indeed the case for ellipticals as well as bulges.
 
Bulges are embedded in large stellar discs in contrast to elliptical galaxies
which poses the question if they really can have the same origin. 
The $\Lambda$CDM paradigm offers a natural way for the transition from 
elliptical galaxies to bulges of early-type spirals via the accretion of a 
new disc in the aftermath of a major merger 
(\cite[Kauffmann et al. 1999]{1999MNRAS.303..188K}; 
\cite[Springel \& Hernquist 2005]{sh05}). As \cite{kb01} show the predicted 
merger rate of galaxies in 
the $\Lambda$CDM paradigm is in fair agreement with the observed one which 
allows to test robustly the transition in Hubble types due to the growth of a 
new stellar disc. Hence the properties of bulges like e.g. the isophotal 
shape (\cite[Khochfar \& Burkert 2005]{kb05}) will initially be set by 
the properties of the progenitor elliptical galaxy.

\section{Model}
We use semi-analytical modelling of galaxy formation to predict the star burst 
and quiescent components of elliptical galaxies. The dark matter history is 
calculated using the merger tree proposed by \cite{som99} with a mass 
resolution of $2 \times 10^9 M_{\odot}$. The baryonic 
physics within these dark matter halos is calculated following recipes 
presented in \cite{kb05} and \cite{2006MNRAS.370..902K}. 
In our simulation, we assume that elliptical galaxies 
form whenever a major merger ($M_1 /M_2 \leq 3.5$ with $M_1 \geq M_2$) takes 
place. We assume that during this process all the cold gas which was in the
 progenitor discs will be consumed in  a central starburst, adding to the 
spheroid mass, and that all stars in the progenitor disks will be 
scattered into the spheroid too. Furthermore we allow the stars of satellite
 galaxies in minor mergers to also contribute to the spheroid.
 During the evolution of a galaxy, we keep track of the origins of all stars 
brought into the spheroid and attribute them to two categories, starburst and 
quiescent, where the first incorporates stars formed during a starburst 
in a major merger and the latter includes stars previously formed in a disc 
and added to the spheroid during a major merger. Each star will 
carry along its 
label and not change it, which means that if a star was made 
in a merger of two progenitor galaxies and the remnant of that merger 
participated in another merger, the star will  still contribute to the 
merger component of the final remnant.  For more modelling details, we refer 
the reader to \cite{2006MNRAS.370..902K} and references therein. Please note 
that our simulation does not include  
AGN-feedback (\cite[Schawinski et al. 2006]{2006Natur.442..888S}) 
 or environmental effects 
(\cite[Khochfar \& Ostriker 2007]{2007arXiv0704.2418K})  that have 
influence on the most massive galaxies.

\section{Results}
\begin{figure}
\center
\includegraphics[height=2.1in,width=2.6in,angle=0]{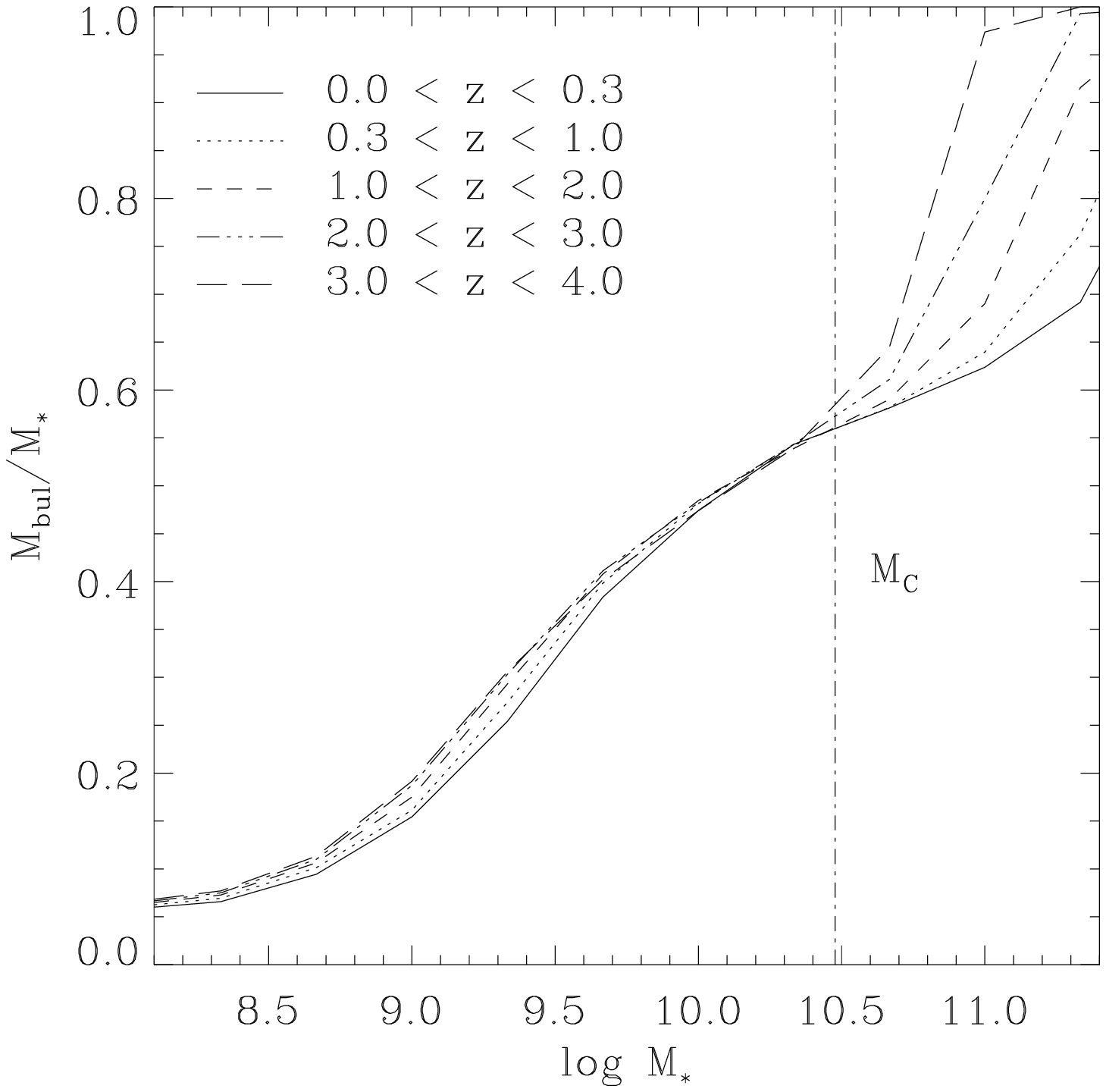}
\includegraphics[height=2.1in,width=2.6in,angle=0]{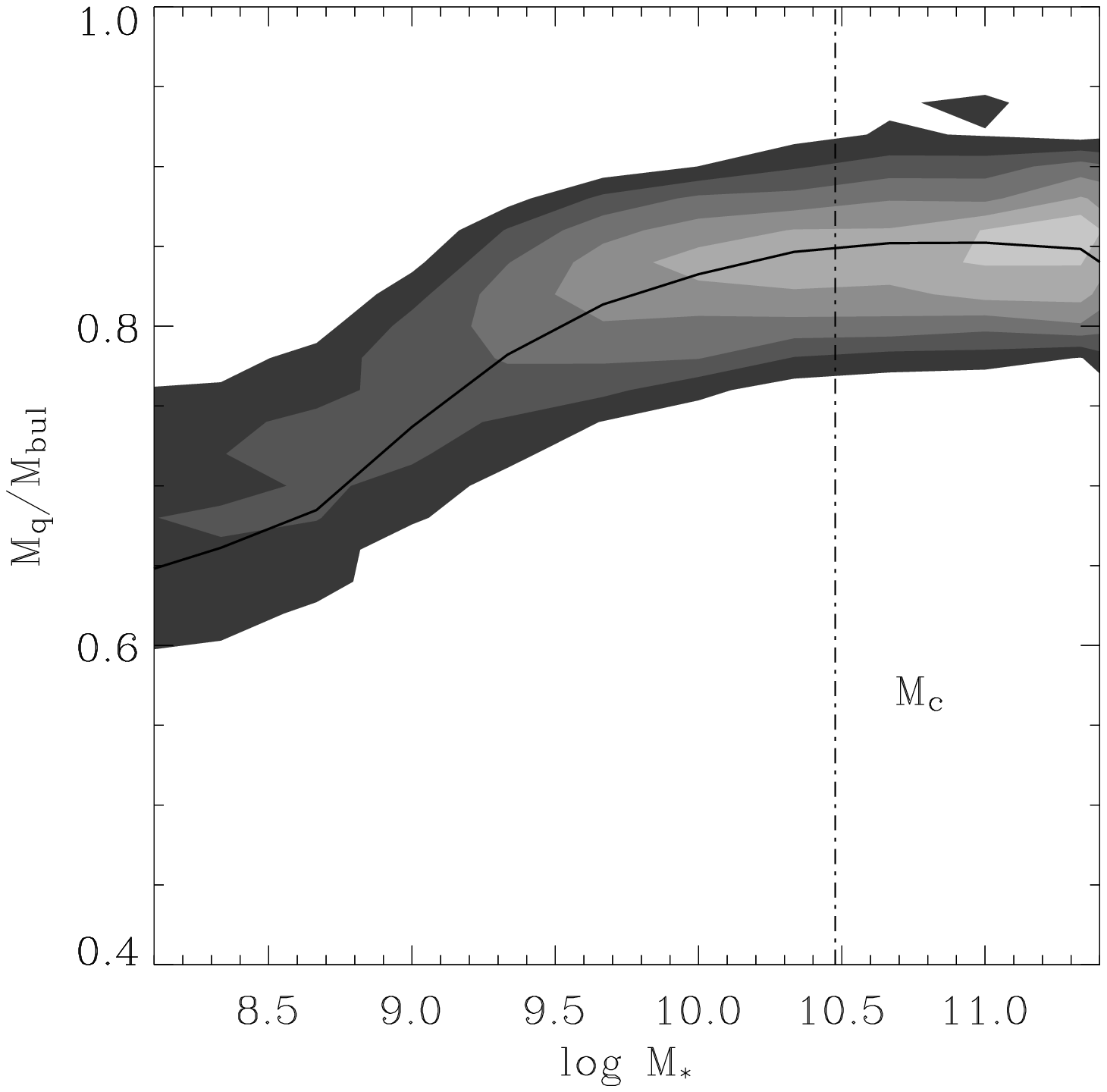}
  \caption{Left panel: Fraction of stars in galaxies of a given mass that 
reside in spheroids at various redshifts. Right panel: Quiescent fraction of 
stars in spheroids as a function of galaxy mass. The solid line shows the 
median of the distribution. The dot-dashed line indicates the critical mass 
scale $M_{\rm{c}}$} \label{f1}
\end{figure}
Ongoing mergers constantly transfer discs stars to spheroids in the universe. 
If this process is more efficient than star formation in discs 
one is to expect an increase in the fraction of stars in spheroids over 
cosmic time. However, the merger rate is a strong decreasing function with 
redshift (\cite[Khochfar \& Burkert 2001]{kb01}) and at late time disc growth 
overtakes merging. In the left panel of figure~\ref{f1} the fraction of 
stars in spheroids 
as a function of redshift and galaxy mass is shown. At early times the most 
massive galaxies, $M_* > M_{\rm{c}}\sim 3 \times 10^{10}$ M$_{\odot}$ 
are all elliptical galaxies and only at late times massive 
spiral galaxies appear. This is related to the gradual transformation of gas 
into stars in discs in contrast to the violent and fast transformation of 
gas into stars during major mergers. Many of the intermediate massive 
elliptical galaxies that formed at high redshift continue to grow discs to 
become bulges of present day spiral galaxies. 

The right panel of figure~\ref{f1} shows the quiescent fraction of bulge stars 
as a function of galaxy mass. The quiescent fraction increases gradually until 
roughly $M_{\rm{c}}$ where it becomes constant at $\sim 0.85$. Most of the 
stars in bulges therefore originated from discs of progenitor galaxies or 
satellite galaxies. \cite{2006MNRAS.370..902K} report the number of 
minor satellite mergers exceeds that of  major mergers by an order of 
magnitude and is therefore one important driver for a high quiescent fraction 
in bulges. For massive bulges in addition mostly dry major mergers cause the 
quiescent fraction to  stay constant and not to change much, which explains 
the behaviour at the high mass end.
 
\begin{figure}
\center
\includegraphics[height=2.15in,width=2.7in,angle=0]{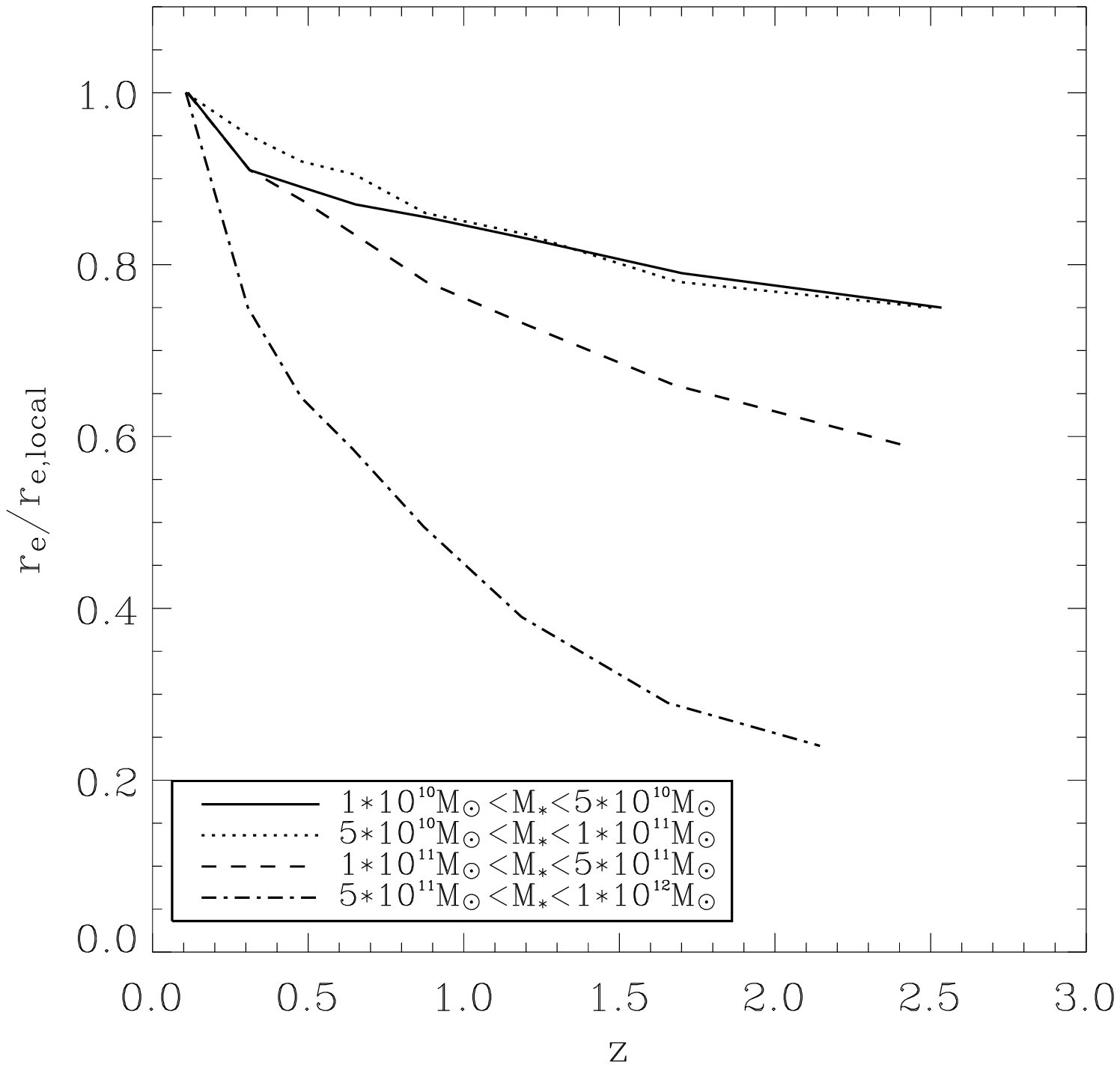}
\includegraphics[height=2in,width=2.5in,angle=0]{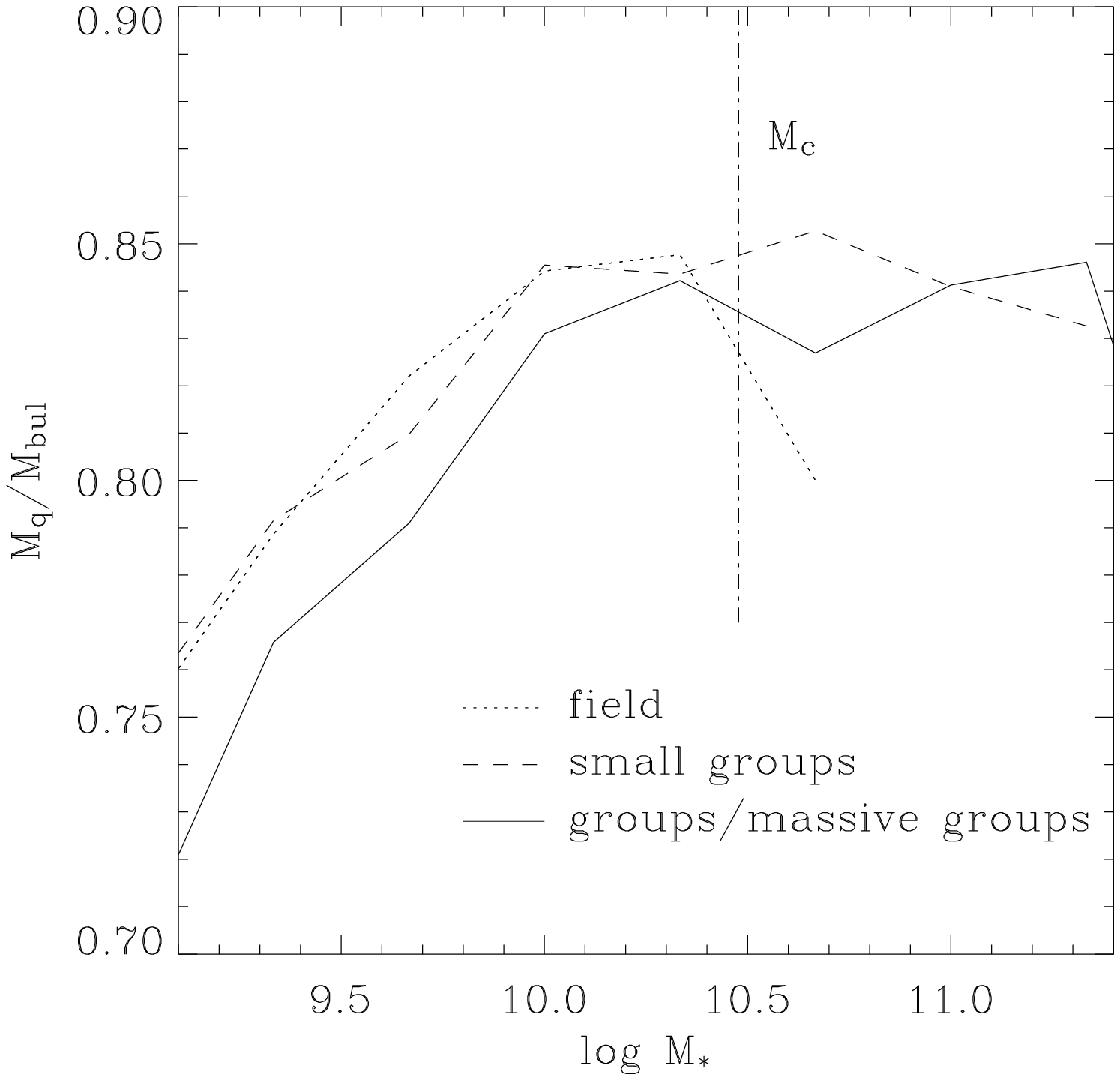}
  \caption{Left panel: size ratio between spheroids of the same mass at high 
    redshift and locally. Right panel: quiescent fraction of stars in 
    bulges as a function of galaxy mass and environment.  }\label{f2}
\end{figure}

Numerical simulations by \cite{sh05} show that dissipation accompanied by 
starbursts during major mergers leads to a population of stars that is more 
centrally concentrated than the scattered disc stars once they relaxed to 
a spheroid at the end of the merger. In our simulations we identified those 
centrally concentrated stars with the starburst component and the less 
concentrated previous disc stars with the quiescent component of bulges. 
\cite{2006ApJ...648L..21K} propose based on these two components a simple 
model in which the size of galaxies scales with the amount of dissipation 
during their formation and that is able to reproduce the size-evolution of 
early-type galaxies. In the left panel of figure~\ref{f2} we show the expected
size evolution of bulges as a function of their mass and formation time, i.e. 
we show the ratio of the present day effective radius of bulges, 
$r_{\rm{e,local}}$, to that of bulges at higher redshifts. 
Massive bulges that formed early are most likely to have had a significant 
amount of dissipation involved during their formation, because 
the gaseous disc only had enough time to transform a small portion of the 
gas into stars. In contrast the size-evolution for small bulges is not very 
strong, as there is only a small difference in the amount of dissipation.

The right panel of the same figure shows the quiescent fraction in bulges as a 
function of galaxy mass and environment. For galaxies more massive than 
$M_{\rm{c}}$ the quiescent fraction does not depend on the environment. 
Only for galaxies  below $M_{\rm{c}}$ we find an environmental dependence 
which reflects itself in a larger quiescent fraction for field galaxies. The 
reason for this is mainly buried in the larger amount of dissipation that is 
involved in the formation of bulges that end up in high density environments. 
These galaxies form in general earlier and therefore  
the amount of dissipation is larger during major mergers.

Observations of core phase-space densities in spiral galaxies 
reveal that they are several order of magnitudes lower than those of 
elliptical galaxies of the same mass 
(\cite[Carlberg 1986]{1986ApJ...310..593C}) . A possible solution to 
this problem is dissipation during starburst that can increase the phase space 
density in the remnant. If the centres of early-type 
spirals are dominated by bulges this suggest that bulges and ellipticals of 
the same mass must have had different amounts of dissipation during their 
formation. Indeed our simulations suggest that the quiescent fraction in 
bulges of spiral galaxies is higher than that of ellipticals of the same mass, 
which could explain the observations.

\section{Conclusions}\label{sec:concl}
Our results presented here indicate that the majority of stars in bulges were
 previously formed in discs and then later added to  bulges by either major or 
minor mergers that occur naturally within the CDM-paradigm. Dissipation 
during mergers however, will be responsible for driving 
the size-evolution and possible environmental dependencies. This work presents 
a first step in differentiating stars in bulges by their origin and it will be 
interesting to see if signature of these two populations can indeed be 
observed.

\end{document}